\documentclass[twocolumn,showpacs,preprintnumbers,amsmath,amssymb]{revtex4}

\usepackage{graphicx}
\usepackage{dcolumn}
\usepackage{bm}

\begin{document}

\preprint{APS/123-QED}

\title{$^{59}$Co NMR evidence for charge ordering below $T_{CO}\sim 51~K$ in Na$_{0.5}$CoO$_2$}

\author{F. L. Ning$^{1}$, S. M. Golin$^{1}$, K. 
Ahilan$^{1}$, T. Imai$^{1,2}$, G.J. Shu$^{3}$, and F. C. Chou$^{3,4}$}

\affiliation{$^{1}$Department of Physics and Astronomy, McMaster University, Hamilton, Ontario L8S 4M1, Canada}
\affiliation{$^{2}$Canadian Institute for Advanced Research, Toronto, Ontario M5G1Z8, Canada}
\affiliation{$^{3}$Center for Condensed Matter Sciences, National Taiwan
University, Taipei 10617, Taiwan}
\affiliation{$^{4}$National Synchrotron Radiation Research Center, HsinChu 30076,Taiwan}

\date{\today}

\begin{abstract}
The CoO$_{2}$ layers in sodium-cobaltates Na$_{x}$CoO$_{2}$ may be viewed as 
a spin $S=\frac{1}{2}$ triangular-lattice doped with charge 
carriers.  The underlying physics of the cobaltates is very similar to that of the high $T_{c}$ cuprates.  We will present 
unequivocal $^{59}$Co NMR evidence that below $T_{CO}\sim51~K$, the insulating ground state 
 of the itinerant antiferromagnet Na$_{0.5}$CoO$_{2}$ ($T_{N}\sim 
86~K$) is induced by charge ordering.\end{abstract}


\maketitle


The discovery of unconventional superconductivity in sodium-cobaltate 
Na$_{1/3}$CoO$_{2}[$H$_{2}$O]$_{4/3}$ ($T_{c}\sim4.5~K$) 
\cite{Takada,Fujimoto,Ishida} has generated major excitement 
in the condensed matter community.  Co ions in the CoO$_{2}$ layers 
take a mixed valence state Co$^{+4-x}$, and form a triangular-lattice.  
Since Co$^{4+}$ and Co$^{3+}$ ions nominally have spins of $S=\frac{1}{2}$ and $S=0$, 
respectively, one may consider the CoO$_{2}$ layers
as a $S=\frac{1}{2}$ triangular-lattice doped with charge 
carriers, in analogy with the doped CuO$_{2}$ 
square-lattice in the high $T_{c}$ cuprate
superconductors.  The fingerprints of Co spins are 
everywhere in Na$_{x}$CoO$_{2}$: Na$_{0.82}$CoO$_{2}$ is an itinerant antiferromagnet 
($T_{N}\sim21~K$) \cite{muSR}; Na$_{0.7}$CoO$_{2}$ is a ``Curie Weiss 
metal'' with large paramagnetic susceptibility \cite{Foo}; Na$_{0.5}$CoO$_{2}$ is an itinerant 
antiferromagnet ($T_{N}\sim 86~K$) \cite{Foo}, and undergoes 
a mysterious metal-insulator transition below $\sim 51~K$ \cite{Foo}.  

 In the case of the cuprates, doped holes tend to segregate themselves from the underlying 
 $S=\frac{1}{2}$ square-lattice to form static {\it charge stripes} \cite{Tranquada}.  
 Whether the charge ordering in stripes is related 
 to or competing against the mechanism of high $T_{c}$ superconductivity is 
 controversial \cite{Millis}.  A major question we address here is whether or not
the cobaltates also exhibit a similar phenomenon in the 
triangular-lattice geometry.  The insulating ground state of 
Na$_{0.5}$CoO$_{2}$ below $\sim 51~K$ has attracted considerable attention 
because of speculation that it may be charge ordered 
\cite{Foo,Balicas,Young,Wang}.  
However, earlier $^{23}$Na NMR measurements in 
Na$_{0.5}$CoO$_{2}$ showed no evidence of the emergence of 
a charge ordered state below $\sim 51~K$ \cite{ETH,Bobroff,NingNa}.  

In this {\it Letter}, we take 
a more direct approach to probing the charge environment in CoO$_{2}$ 
layers, by measuring the EFG ({\it Electric Field Gradient}) of Co sites with zero-field $^{59}$Co NMR.  The EFG tensor is the second derivative of the Coulomb potential, and hence is directly related to the local charge density.   
We will present unequivocal evidence that  the insulating ground 
state of Na$_{0.5}$CoO$_{2}$ below $T_{CO}\sim 51~K$ 
is indeed the consequence of charge ordering.
 
A major distinction between cobaltates and cuprates is that 
the Na$^{+}$ ions in cobaltates can spatially order for certain values 
of Na concentration, $x$.  Besides donating electrons, 
ordered Na$^{+}$ ions exert a periodic Coulomb 
potential on the CoO$_{2}$ triangular-lattice 
\cite{Foo,Zandbergen,Huang,Zhang,Roger,Chou}.   
In the case of Na$_{0.5}$CoO$_{2}$, electron and neutron diffraction measurements  
 have suggested that Na$^{+}$ ions form zigzag chains \cite{Foo,Zandbergen,Huang}.  
 This results in the presence of 
two structurally inequivalent Co sites, each forming a chain 
structure\cite{Huang} as shown in 
Fig.\ 1.  One of the nearest neighbor sites of Co(1) is occupied by 
a Na$^{+}$ ion, while Co(2) has no Na$^{+}$ ions in the nearest 
neighbor sites \cite{Huang}.  
Since the Coulomb potential from 
 Na$^{+}$ chains attracts electrons, the valence of Co(1) sites,
 Co$^{+3.5-\delta}$, is smaller than that of Co(2) sites, 
 Co$^{+3.5+\delta}$.  The average valence of Co ions in Na$_{0.5}$CoO$_{2}$ is 
 +3.5. The valence of Co(2) sites is closer to Co$^{+4}$ 
($S=\frac{1}{2}$), hence Co(2) exhibits strong spin fluctuations above 
$T_{N}\cong 86$~K \cite{Ning}, and develops a sizable ordered magnetic
moment $\sim 0.26 \mu_{B}$ within the CoO$_{2}$ plane below $T_{N}$
\cite{Yokoi,Young}.  On the other hand, Co(1) sites have much
weaker spin fluctuations above $T_{N}$\cite{Ning}.  
The upper bound of the ordered moment at Co(1) 
sites is as small as $\sim 0.04 \mu_{B}$ at 8 K and too 
small to be detected by polarized neutron scattering techniques \cite{Young}.  
$^{59}$Co zero field NMR is a very powerful technique for detecting 
small hyperfine magnetic field $B_{hf}$, which is proportional to the 
ordered magnetic
moments.  Yokoi et al. took advantage of this high sensitivity, and demonstrated the presence of 
small ordered  moments at Co(1) sites along the crystal c-axis below $T_{N}\sim 86~K$.  This led them to propose  
two possible spin structures  as shown in Fig.1 \cite{Yokoi}.

Although the existence of multiple Co 
sites in Na$_{x}$CoO$_{2}$ is sometimes referred to as a consequence of  
{\it charge ordering} persisting up to room temperature, 
it is important to realize that these distinct 
behaviors of Co(1) and Co(2) sites in Na$_{0.5}$CoO$_{2}$ are 
directly linked with the periodic Coulomb potential arising from Na$^{+}$ 
zigzag chains.  Moreover, charges on the Co sites are 
mobile, and Na$_{0.5}$CoO$_{2}$ is metallic 
above $T_{CO}\sim 51$~K \cite{Foo}.  
In contrast, strong electron-electron 
correlation effects in high $T_{c}$ cuprates induce a
self-organized pattern of doped carriers in the form of charge stripes  
\cite{Tranquada}.  The 
spatial periodicity of the charge ordered state is different from that of the Coulomb 
potential of the underlying lattice.  It is this type of 
self-organized charge pattern that we are searching for in Na$_{0.5}$CoO$_{2}$.

In Fig.2, we present typical zero-field $^{59}$Co (nuclear spin $I=\frac{7}{2}$) 
NMR lineshapes 
from a piece of electrochemically 
deintercalated Na$_{0.5}$CoO$_{2}$ single crystal (mass $\sim
25$mg) \cite{Chou2,Chou3}. 
The observed lineshapes are similar to Yokoi's\cite{Yokoi}.  
Below $T_{N}\sim 86~K$, we observe 7 NMR 
peaks of Co(2) sites for transitions between nuclear spin $I_{z}=\frac{2m+1}{2}$
and $I_{z}=\frac{2m-1}{2}$, where integer $-3 \leq m \leq +3$.  
In the frequency range above 5~MHz, we have also
successfully detected Co(1) zero-field NMR signals for the $m = 0$ to 
$m=+3$ transitions.  The intensity of NMR signals decreases in proportion to 
the square of the frequency, and measurements below 5~MHz are formidable.  
We summarize the temperature dependence of the observed peak 
frequencies $f_{m}$ of Co(1) sites in Fig.3.
A remarkable feature of Fig.2 and Fig.3 is that 
all 4 peaks of Co(1) sites split into
two precisely below $T_{CO}\sim 51$~K.  In contrast, Co(2) sites 
show no anomalies.  In what follows, 
we will refer to the two inequivalent Co(1) sites as Co(1a) and Co(1b). 
In passing, the comparison between the zero-field and low-field NMR 
lineshapes indicates that the NMR signals near
6~MHz, shown in green, are not related to Co(1) sites 
\cite{NingNa}.  We tentatively attribute these
unidentified signals to defect sites caused by slight
deviation of Na concentration from $0.5$ and/or minor disorder.   We also confirmed that both Co(1a) and Co(1b) lines split in weak magnetic field applied along the c-axis \cite{NingNa}.  This means that both Co(1a) and Co(1b) sites have up-spins as well as down-spins.

To understand the effects of the EFG on NMR lineshapes, we need to 
theoretically compute the resonance frequency $f_{m}$ of the $m$-th transition as $f_{m}
=(E_{(2m+1)/2}-E_{(2m-1)/2})/h$,
where the energy levels
$E_{n}$ are eigenvalues of the standard $^{59}$Co nuclear spin Hamiltonian,
\begin{equation} \label{1}
\ H =\frac{\textit{h} \nu_Q^{Z}}{6}\{3I_{Z}^2- I(I+1) +\eta(I_{X}^2-I_{Y}^2)\} -\gamma_n\textit{h}\textbf{B} \cdot \textbf{I}.
\end{equation}
The first term of the Hamiltonian represents the nuclear quadrupole
interaction \cite{Das} between the EFG 
and the $^{59}$Co nuclear quadrupole moment.
The diagonalized tensor of the nuclear quadrupole interaction ($\nu_Q^{X}$, $\nu_Q^{Y}$, $\nu_Q^{Z}$) is
proportional to the EFG tensor ($d\phi^2/d X^2$, $d\phi^2/d Y^2$, $d\phi^2/d Z^2$), where $\phi$ is the total Coulomb potential seen by $^{59}$Co nuclei, and $X$, $Y$ and $Z$ represent three orthogonal,
principal axes (by following the convention, we define
$|\nu_Q^{X}| < |\nu_Q^{Y}|< |\nu_Q^{Z}|$). From the
comparison of zero-field NQR (Nuclear Quadrupole
Resonance) results at 110~K (see Fig.2) and  high-field NMR, we found that the main-principal axis
$Z$ coincides with the crystal c-axis within experimental 
uncertainties ($\sim 5$~degrees).  This
means that the other principal axes, $X$ and $Y$, lie within the CoO$_{2}$
plane.  Poisson's relation for the Coulomb interaction sets a
constraint, $\nu_Q^{X}+\nu_Q^{Y}+\nu_Q^{Z}=0$.  Hence we have only
two independent parameters in the first term of eq.(1), $\nu_Q^{c}$ (we now denote $Z=c$) and
the anisotropy parameter $\eta = (\nu_Q^{Y} -\nu_Q^{X})/\nu_Q^{Z}$. $\eta$ is a measure of the
deviation from axial symmetry of the EFG tensor with respect to the
crystal c-axis.  In the paramagnetic state at 110~K ($> T_{N}$), we deduced from the $^{59}$Co
NQR lineshape  that $\nu_{Q}^{c}=2.799(5)$ MHz and 
$\eta= 0.300(1)$ for Co(1)
sites, and $\nu_{Q}^{c}=4.040(5)$
MHz and $\eta= 0.400(1)$ for Co(2) sites. 

The second term in the Hamiltonian represents the
Zeeman interaction between Co nuclear spins and the
 {\it local} magnetic field, {\bf {B}}, at the position of the observed nuclear
 spin.  $\gamma_n = 2\pi\times 10.054$
MHz/Tesla is the $^{59}$Co nuclear gyromagnetic ratio.  
In our zero-field NMR measurements, we apply no external 
magnetic field.  Hence ${\bf B} ={{\bf B}_{hf}}$, where ${{\bf B}_{hf}}$
is the hyperfine magnetic field from ordered Co moments. 
 ${\bf B}_{hf}$ is proportional to the sublattice magnetization, 
 ${\bf M}(T)$, of Co 
 sites \cite{Yokoi}.  The $m=0$ transition
 frequency depends primarily on the Zeeman term, $f_{0}\sim \gamma_n
 |{\bf B}_{hf}|$.  At 60~K, $f_{0} = 6.659$ MHz and
$19.780$ MHz for Co(1) and Co(2) sites, respectively, because
the hyperfine magnetic field is ${|{\bf B}_{hf}|}\sim 0.6$ Tesla
at Co(1) sites, and $\sim 1.9$ Tesla at Co(2) sites.  The
separation between adjacent peaks, $(f_{m+1}-f_{m})$, depends
primarily on the EFG.  For Co(1) sites, $(f_{m+1}-f_{m})\sim\nu_{Q}^{c} 
\sim 2.8$ MHz, since the ordered moments point along the c-axis. 
For Co(2) sites, the separation is $(f_{m+1}-f_{m})\sim\nu_{Q}^{a,b} \sim\nu_{Q}^{c}/2 \sim 2$
MHz or less, because the ordered magnetic moments point within the
ab-plane.  The observed peaks $f_{m}$ are not evenly 
spaced, because the nuclear quadrupole and Zeeman 
interactions are comparable in the present case.   This means that 
we must diagonalize the Hamiltonian in eq.(1) exactly to fit the 
lineshapes.  By relying on inaccurate second-order perturbation analysis, Yokoi et 
al. not only misidentified the peaks of Co(2) sites but were also 
unable to deduce the critically important information associated with the Co(1) sites \cite{Yokoi}. 
    
Based on the exact diagonalization, we matched all the observed zero-field NMR peaks
with theoretically calculated values of $f_{m}$ as shown by vertical lines in Fig.2.  This allowed us to deduce the relevant NMR
parameters of eq.(1), {\it i.e.} $\nu_{Q}^{c}$,
$\eta$, the magnitude of $B_{hf}$, and the relative polar angle ($\theta$, $\phi$) between
$B_{hf}$ and the c-axis.  We found that $\eta$ shows no temperature dependence across 51~K, 
and ($\theta$, $\phi$) change
less than 2 degrees from the values at 60~K ($\theta=9\pm 1^{\circ}$, $\phi=0^{\circ}$).  
Hence we focus our attention on the temperature dependence of $\nu_{Q}^{c}$ and $B_{hf}$
 of Co(1a) and Co(1b) sites.  

Co(1a) and Co(1b) have different values of $\nu_{Q}^{c}$, as shown in Fig4b.  This is direct proof
that local charge environment is different between Co(1a) and Co(1b) 
sites.  In principle, a differentiation of the EFG between Co(1a) and Co(1b) sites could arise if a structural 
phase transition doubles the unit cell of Na$^{+}$ zigzag chains.  
To rule out such a scenario, we deduced $\nu_{Q}^{c}(Na)$ at two structurally inequivalent Na sites, Na(1) and Na(2) \cite{Huang}, from the measurements of $+\frac{1}{2}$ to $-\frac{1}{2}$ central transition and $\pm\frac{3}{2}$ to $\pm\frac{1}{2}$ satellite transitions of $^{23}$Na NMR in the same crystal \cite{NingNa}.  As summarized in Fig.4a, we found no hint of additional splitting or extra broadening for $\nu_{Q}^{c}(Na)$ below $T_{CO}$.  The upper bound of the potential splitting of $\nu_{Q}^{c}(Na)$ at 4~K is less than $\sim 1\%$, while the splitting of  $\nu_{Q}^{c}$ between  Co(1a) and Co(1b) sites, $\Delta \nu_{Q}^{c} = \nu_{Q}^{c,~Co(1b)}-\nu_{Q}^{c,~Co(1a)}$, reaches as much as $\sim 6\%$, as shown in Fig.4d.  Therefore we conclude that {\it Co(1) sites undergo a charge ordering transition at $T_{CO}\sim$51~K}.

The observed differentiation of $B_{hf}$ between Co(1a) and Co(1b) sites in 
Fig.4c is also understandable as a consequence of charge ordering.  Since 
the valence state is different below $T_{CO}\sim 51~K$, the number of electrons filling the 
$t_{2g}$ orbitals of Co(1a) and Co(1b) sites will also be different.  This 
naturally results in a different magnitude of ordered moments at Co(1a) and Co(1b) sites, 
and hence separate values of $B_{hf}$.  In Fig.4d, we plot the splitting of $B_{hf}$ between Co(1a) and Co(1b) sites, $\Delta B_{hf} = B_{hf}^{Co(1a)} - B_{hf}^{Co(1b)}$.  $\Delta B_{hf}$ shows an identical temperature dependence as $\Delta \nu_{Q}^{c}$ within experimental
uncertainties.  This is not surprising, because to a good approximation, both $\Delta \nu_{Q}^{c}$ and $\Delta B_{hf}$ should be proportional to the difference of the number of electrons occupying the Co 3d orbitals.  Thus we may consider  $\Delta \nu_{Q}^{c}$ and  $\Delta B_{hf}$ as
{\it the order parameter} of charge ordering below 51~K.  By fitting the temperature 
dependence down to 40~K ($= 0.8 T_{CO}$) to a
typical power law behavior,  $\Delta \nu_{Q}^{c} \sim \Delta B_{hf} \sim (T-T_{CO})^{\beta'}$, we found the
critical exponent $\beta' = 0.3 \pm 0.1$.  Interestingly, this value  is comparable to the critical exponent $\beta = 0.28 \pm 0.02$ observed for the sub-lattice magnetization, $M(T) \sim (T-T_{N})^{\beta}$ at $T_{N}=86$ K for the antiferromagnetic N{\'e}el transition of Co(2) sites \cite{Young}.   

To conclude, we have presented unequivocal zero-field NMR evidence 
that static differentiation of charge densities develops below 
$T_{CO}\sim 51$~K 
in the one dimensional chain structure of Co(1)
sites.  In contrast,  
Co(2) sites have only spin ordering and show no anomalies either in NMR lineshapes or $\nu_Q$ across $51$~K.
In Fig.1, we show the possible charge ordering patterns with the 
smallest possible unit cell.    
Our observation of a striped configuration of the  charge ordered 
state, consisting of Co(1) chains with charge and spin ordering and 
Co(2) chains with spin ordering only, is consistent with the two-fold 
symmetry observed for in-plane magnetoresistance by Balicas et al. 
\cite{Balicas}.  The splitting of $\nu_{Q}^{c}$ (and hence EFG)
reaches as much as $\sim 6$~\% between Co(1a) and 
Co(1b) sites.    We recall that  $\nu_{Q}^{c}$ at $^{63}$Cu sites 
increases by $\sim 10$\% in high $T_c	$ cuprates when the hole
concentration $x$ changes by 0.15 in the CuO$_{2}$ plane of La$_{2-x}$Sr$_{x}$CuO$_{4}$ \cite{imai93}.  
If we assume that the charge doping effects on the EFG are comparable 
between cobaltates and cuprates, we can crudely estimate the 
differentiation of the valence between Co(1a) and Co(1b) 
sites as $\sim 0.1$.  It is worth noting that t-U-V 
model calculations \cite{Wang} showed that {\it a differentiation of 
valence of Co(2) sites} as little as $\sim 
0.03$ may be sufficient to drive CoO$_2$ planes into insulating.
On the other hand, none of the charge ordered patterns proposed within the existing theoretical frameworks \cite{Wang,Ogata,Phillips} are consistent with our experimental finding of charge ordering {\it on Co(1) sites}.  Clearly, more theoretical works are required.   \\

{\bf Acknowledgment}\\
We thank P.A. Lee for helpful communications and encouragement.  
TI acknowledges support from NSERC and CIFAR. FCC acknowledges
support from NSC-Taiwan under contract number NSC-95-2112-M-002.\\


 \vspace {0.5in}

Fig.1.  Ordered magnetic moments on Co(1) sites point either $\textit{up}$ 
($\bullet$) or $\textit{down}$ ($\times$), while moments on Co(2) 
sites point within 
the plane (arrows) \cite{Yokoi}.  Co(1) sites have either in-plane 
antiferromagnetic order (left panel) or in-plane ferromagnetic order (right 
panel).  In the latter case, Co(1) sites in the adjacent CoO$_{2}$ layers 
have opposite spin orientation.  Dark blue and light blue on Co(1a) and Co(1b) sites, respectively, represent possible charge ordering patterns to be determined in this work.  Red dashed lines represent the unit 
cell in each possible configuration.\\

Fig.2.  $^{59}$Co NMR lines at Co(1) sites (blue) and Co(2) sites (orange) 
at 110 K (paramagnetic state, $B_{hf}=0$), 60 K (N{\'e}el ordered state), 
30 K and 4.2 K (charge ordered state). 
Vertical lines represent theoretical fit of resonance frequencies 
$f_{m}$ for  Co(1) (black), Co(1a) (dark blue), Co(1b) (light blue), 
and Co(2) (orange). 
Peak(s) near 6~MHz (green) are not associated with Co(1) or Co(2) 
sites (see main text). \\

Fig.3.  $f_{m}$ ($m= 0, 1, 2$ and $3$) for Co(1) (black) above 
$T_{CO}$, and Co(1a) (dark blue) and Co(1b) (light blue) below $T_{CO}$. 
Vertical arrows roughly represent the magnitude of
$\nu_{Q}^{c}$ for Co(1a) and Co(1b) sites.\\

Fig.4.  (a) Temperature dependence of $\nu_{Q}^{c}(Na)$ at Na(1) and Na(2) sites.  The distribution of $\nu_{Q}^{c}(Na)$ is about the size of the symbols.  (b) Temperature dependence of $\nu_{Q}^{c}$, and 
(c) hyperfine field $B_{hf}$ in Co(1) sites.  Color convention is the same as in 
Fig.2 and Fig.3.  (d) $\Delta\nu_{Q}^{c}$ : the
splitting of $\nu_{Q}^{c}$ between Co(1b) and Co(1a) sites deduced from Fig.4b (left axis, diamond).  $\Delta B_{hf}$ : the splitting of $B_{hf}$ between Co(1a) and Co(1b) deduced from Fig.4c (right axis, filled circles). Solid 
curve shows a best fit to $\Delta\nu_{Q}^{c} \sim  \Delta B_{hf} \sim (T-T_{CO})^{\beta'}$ with a critical exponent $\beta'=0.3$.  The dashed line shows T$_{CO} \sim 51~K$. \\

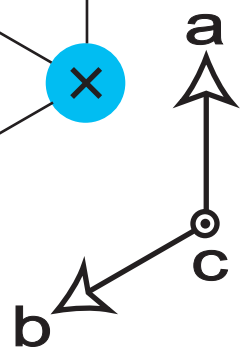
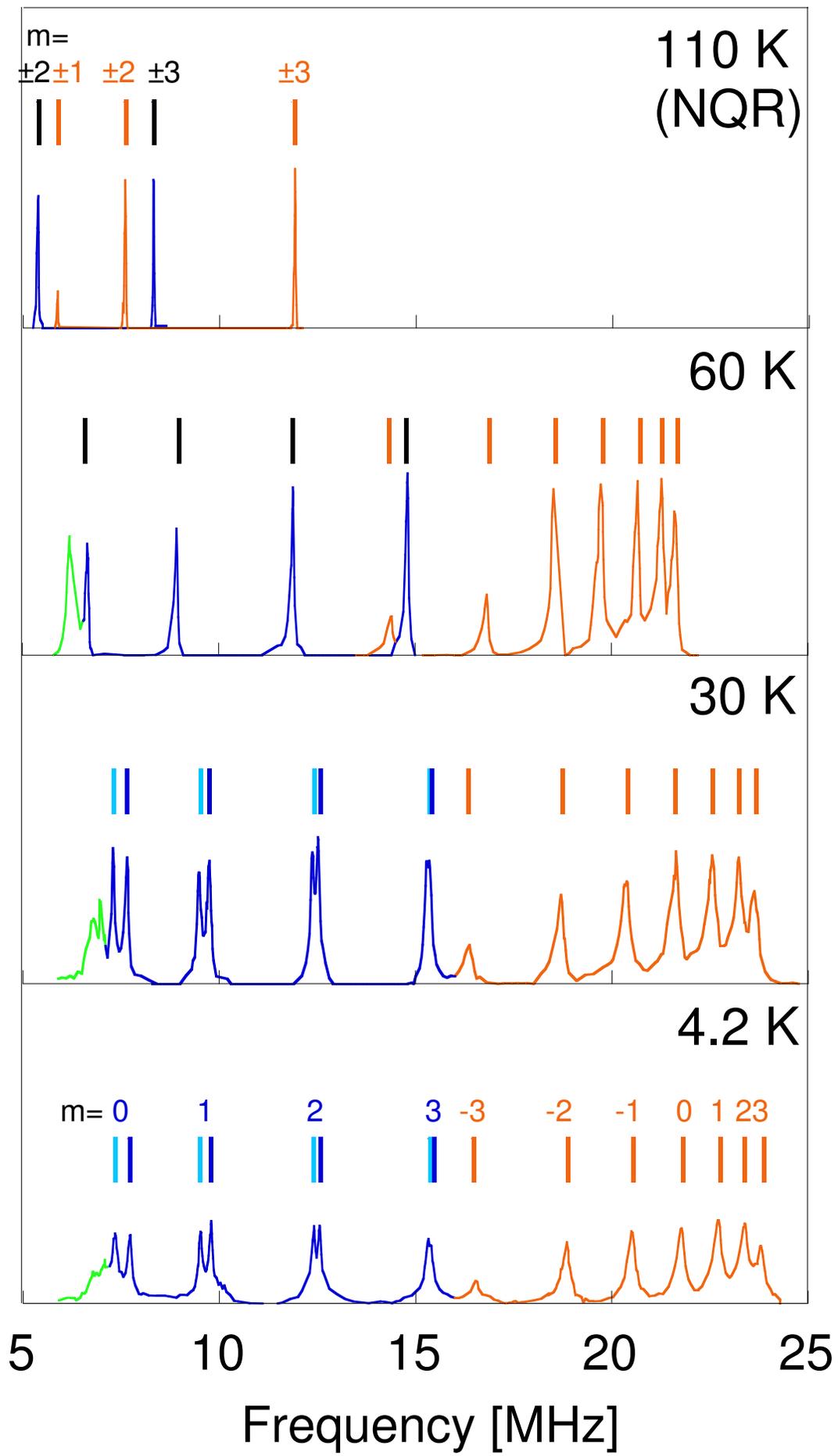
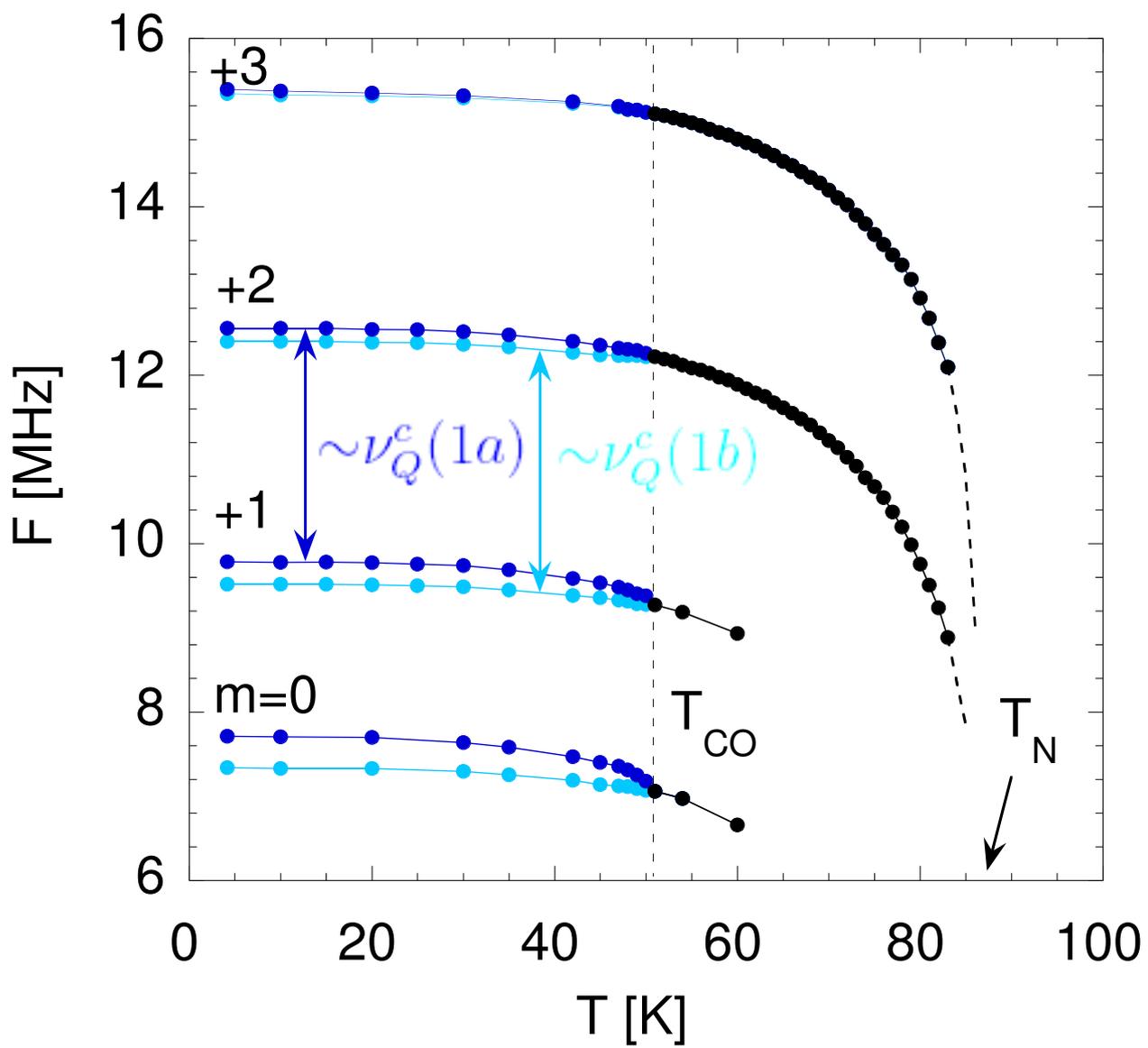
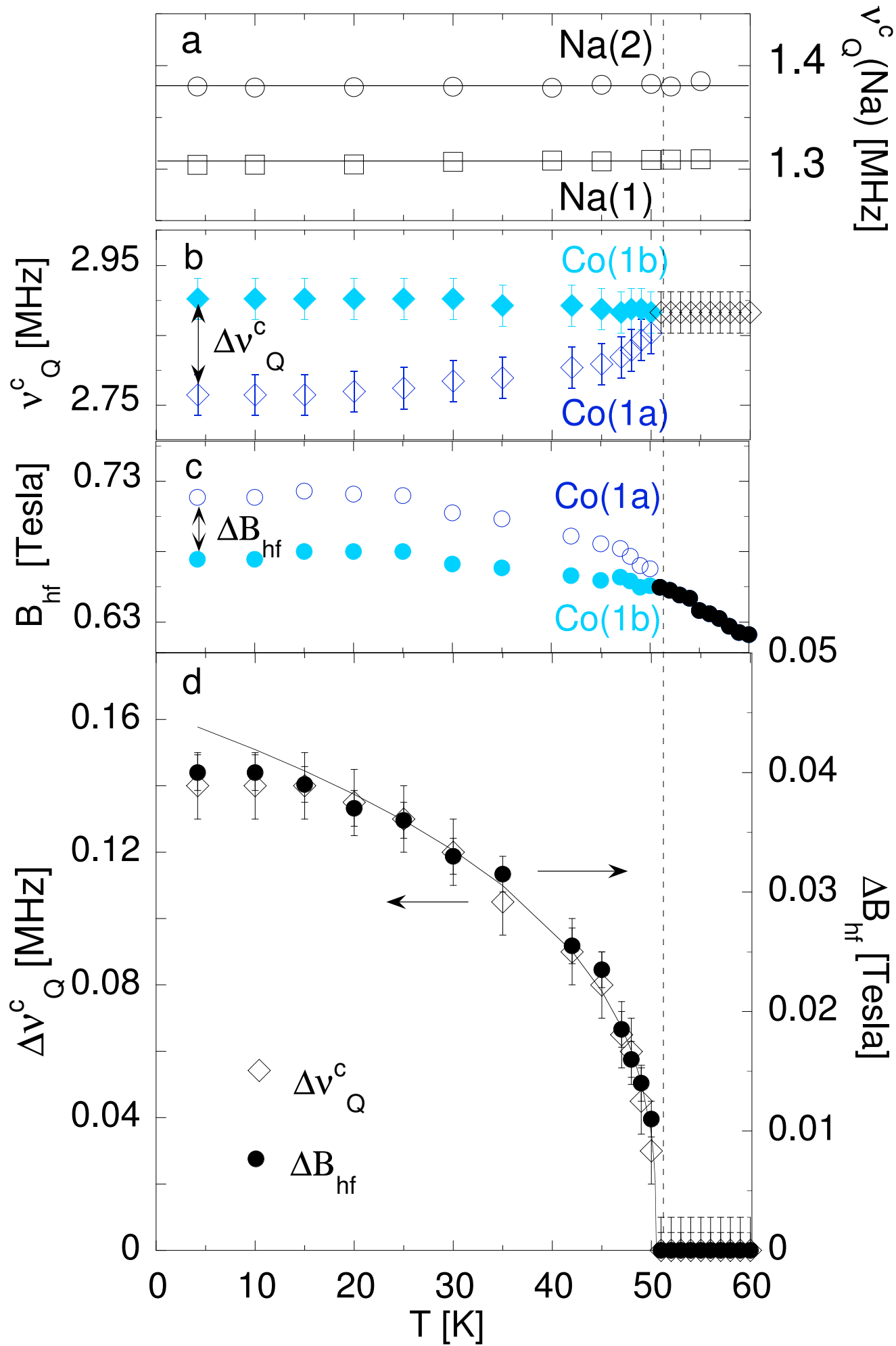
\end{document}